\documentclass{webofc}
\usepackage{feynmp-auto}
\usepackage[varg]{txfonts}
\begin{document}
\title{Modifying the theory of gravity by changing independent variables
}
\author{\firstname{N.V.} \lastname{Kharuk}\inst{1}\fnsep\thanks{\email{natakharuk@mail.ru}}
\firstname{S.N.} \lastname{Manida}\inst{1}\fnsep\thanks{\email{s.manida@spbu.ru}}
\firstname{S.A.} \lastname{Paston}\inst{1}\fnsep\thanks{\email{s.paston@spbu.ru}}
\firstname{A.A.} \lastname{Sheykin}\inst{1}\fnsep\thanks{\email{a.sheykin@spbu.ru}}}

\institute{ Saint Petersburg State University, Saint-Petersburg, Russia}

\abstract{ We study some particular modifications of gravity in search for a natural way to unify the gravitational and electromagnetic interaction. The certain components of 
connection in the appearing variants of the theory can be  identified with electromagnetic potential.
The methods of adding matter in the form of 
scalar and spinor fields are studied.
In particular, the expansion of the local symmetry group up to
$GL(2,C)$ is explored, in which equations of Einstein,
Maxwell and Dirac are reproduced for the theory with
Weyl spinor.
}
\maketitle
\section{Introduction}
The gravity have been described by General Relativity (GR) for more than hundred years. In spite of this, various modifications of this theory are still being discussed. One of the reason for this is the inability to construct a correct quantum theory of gravity. In the classical theory, within the framework of the modified theory, one can try to solve problems of dark energy, dark matter and the problem of energy.  Another  aim of modification is the construction of the unified theory of gravity and other interactions.

The alternative description of GR in which metric and connection are considered as independent variables is well studied, see for example  \cite{1}. It is called Palatini formalism. Usually this approach is equivalent to the GR. More interesting results are obtained if one considers Palatini formalism with one more generalization, namely nonsymmetric connection. This approach we will discuss in this paper. For the first time such approach with symmetric metric was proposed only in the second half of the twentieth century \cite{2}, \cite{3}.  It turns out that this theory can be interpreted as unified theory of gravity and electromagnetism, however, some additional assumptions were used by the authors to introduce a matter in the theory. A more natural way  to add a classical matter in such theory was proposed in \cite{4}, \cite{5}.

In present work the study of Palatini formalism with nonsymmetric connection is continued. Addition of matter in the form of scalar and spinor field are described.  The tetrad formalism is used to describe spinor field and  the expansion of the local symmetry group up to
$GL(2,C)$ is studied.

\section{Palatini formalism with nonsymmetric connection}
First,  let us consider the Einstein-Hilbert action without matter in a standard form:
\begin{equation}
S_1=-\frac{1}{2\varkappa} \int d^4x \sqrt{-g} g^{\mu\nu} R_{\mu\nu}(\Gamma),
\end{equation}
where $\varkappa$ is the Einstein gravitational constant, $R_{\mu\nu}$ is Ricci tensor, $g \equiv \text{det } g_{\mu\nu}$. The main differences between this action and the standard one are that the Ricci tensor depends only on the connection and  connection is an independent nonsymmetric variable. For the first time such action was considered by Einstein \cite{5}, but he also assumed that the metric is nonsymmetric too.

It is easy to show that the theory with action (1) is completely equivalent to the GR. To do it let us introduce the new quantity $\hat{\Gamma}^\alpha_{\mu\nu}$, which is equal to the difference between the independent connection and Christoffel symbols $\bar{\Gamma}^\alpha_{\mu\nu}(g)$, which are  expressed in a standard way through the metric:
\begin{equation}
\hat{\Gamma}^\alpha_{\mu\nu}=\Gamma^\alpha_{\mu\nu}-\bar{\Gamma}^\alpha_{\mu\nu}(g).
\end{equation}
Using this relation, one can write the expression for a curvature tensor in such form:
\begin{equation}
    R^\mu{}_{\nu\alpha\beta}=\bar{R}^\mu{}_{\nu\alpha\beta}(g)+\bar{D}_\alpha \hat{\Gamma}^\mu_{\beta\nu}-\bar{D}_\beta \hat{\Gamma}^\mu_{\alpha\nu}
 +\hat{\Gamma}^\mu_{\alpha\xi}\hat{\Gamma}^\xi_{\beta\nu}-\hat{\Gamma}^\mu_{\beta\xi}\hat{\Gamma}^\xi_{\alpha\nu},
\end{equation}
where $\bar{R}^\mu{}_{\nu\alpha\beta}(g)$ is a Riemann tensor, which is expressed in a standard form using the Christoffel symbols, and covariant derivative $\bar{D}_\beta$ includes  the Christoffel symbols only.
Then the  action (1) can be written as (omitting the surface terms)
\begin{multline}
    S_1=-\frac{1}{2\varkappa} \int d^4x \sqrt{-g} ( g^{\mu\nu} \bar{R}_{\mu\nu}(g)+\hat{\Gamma}^\mu_{\mu\xi}\hat{\Gamma}^\xi_{\beta\alpha}g^{\beta\alpha}-
 \hat{\Gamma}^\mu_{\beta\xi}\hat{\Gamma}^\xi_{\mu\alpha}g^{\alpha\beta})=\\=
 \bar{S}_1-\frac{1}{4\varkappa} \int d^4x \sqrt{-g}
 \hat{\Gamma}^\mu_{\alpha\gamma}\hat{\Gamma}^\delta_{\beta\nu} L_\mu{}^{\alpha\gamma,}{}_\delta{}^{\beta\nu},
\end{multline}
where 
$\bar{R}_{\mu\nu}(g) \equiv \bar{R}^\alpha{}_{\mu\alpha\nu}(g) $ and 
\begin{equation}
L_\mu{}^{\alpha\gamma,}{}_\delta{}^{\beta\nu}=\delta^\gamma_\delta \delta^\alpha_\mu g^{\beta\nu}-
\delta^\gamma_\delta \delta^\beta_\mu g^{\alpha\nu}+\delta^\beta_\delta \delta^\nu_\mu g^{\alpha\gamma}-
\delta^\alpha_\delta \delta^\nu_\mu g^{\gamma\beta}
\end{equation}
is a $64\times 64$ matrix. It has the symmetry property: $L^{IK}=L^{KI}$, where multindex $I\equiv \left\{{}_\mu{}^{\alpha\gamma}\right\}$ and it is easy to verify that this matrix has 4-dimensional zero proper subspace: $L_\mu{}^{\alpha\mu,}{}_\delta{}^{\beta\nu}=0$.

Let us split $\hat{\Gamma}$ to the traceless and trace parts:
 \begin{equation}
  \hat{\Gamma}^\alpha_{\mu\nu}=\underbrace{\tilde{\Gamma}^\alpha_{\mu\nu}}_{\tilde{\Gamma}^\alpha_{\mu\alpha}=0}+
  \underbrace{\frac{1}{4}\hat{B}_\mu \delta^\alpha_\nu}_{\hat{B}_\mu=\hat{\Gamma}^\alpha_{\mu\alpha}}.
 \end{equation}
Using this notations one can express the action $S_1$ in the following form:
\begin{equation}
  S_1=-\frac{1}{2\varkappa} \int d^4x \sqrt{-g}  g^{\mu\nu} \bar{R}_{\mu\nu}(g)-
  \frac{1}{4\varkappa} \int d^4x \sqrt{-g}
 \tilde{\Gamma}^\mu_{\alpha\gamma}\tilde{\Gamma}^\delta_{\beta\nu} L_\mu{}^{\alpha\gamma,}{}_\delta{}^{\beta\nu}.
\end{equation}
The first term is a standard Einstein-Hilbert action, which depends only on the metric, and the second one is an additional contribution. However, it is easy to calculate that the last term does not change the equations of motion at all, because after variation with respect to $\tilde{\Gamma}^\mu_{\alpha\gamma}$ one can obtain $\tilde{\Gamma}^\mu_{\alpha\gamma}=0$. Thus after variation we get the standard Einstein equations. It means that this theory reproduces the GR. 

\section{Addition of matter}
\subsection{A classic point-like particle}
The most simple case of matter is a classic point-like particle. This variant was described in detail in \cite{4}, here we give briefly the results. Classical particle  is described by standard action in 
a curved space-time:
\begin{equation}
S_m= -m\int d\tau \sqrt{\dot{x}^\mu(\tau) \dot{x}^\nu(\tau) g_{\mu \nu}(x(\tau))}   , 
\end{equation}
where $m$ is a mass of particle and $x^\mu(\tau)$ describes the world line of the particle.
The interaction of classical particle with the connection is introduced to get a more general theory. One of the  simplest ways is 
\begin{equation}
 S_{int}=-q \int d\tau \dot{x}^\mu(\tau) \hat{B}_\mu(x(\tau)),
\end{equation}
where $q$ is a parameter. 
It turns out that to get a self-consistent theory we need to add one more term to the total action. It is the kinetic term which can be written in different ways, but again the most simple case is chosen

\begin{equation}
   S_2= -\frac{1}{16\pi} \int d^4x \sqrt{-g} R^\mu{}_{\mu\alpha\beta}R^\nu{}_{\nu\rho\sigma}g^{\alpha\rho}g^{\beta\sigma},
\end{equation}
where contraction of curvature $R^\mu{}_{\mu\alpha\beta}$ is so-called segmental curvature, which depends only on $\hat{B}_\mu$ (using (3)):
\begin{equation}
R^\mu{}_{\mu\alpha\beta}=\partial_\alpha \hat{B}_\beta-\partial_\beta \hat{B}_\alpha.
\end{equation}

As a result, the total action consists of four terms: $S=S_1+S_m+S_{int}+S_2$.  It depends on three independent variables: the metric $g_{\mu\nu}$,  the coordinate of the particle $x^\mu(\tau)$ and 
 the connection $\Gamma^\mu_{\alpha\gamma}$. The full  system of motion equations is  the following:
  \begin{equation}
  \begin{cases}
   \tilde{\Gamma}^\alpha_{\mu\nu}=0;\\
    \bar{R}_{\mu\nu}-\frac{1}{2}\bar{R}g_{\mu\nu}=\varkappa (T_{1\mu \nu} + T_{2\mu\nu});\\
     \bar{D}_\mu F^{\mu\nu}=4\pi j^\nu;\\
  m u^\mu \bar{D}_\mu u^\alpha= -qu_\xi F^{\xi\alpha},
  \end{cases}
\end{equation} 
where  $F_{\mu\nu}\equiv R^\alpha{}_{\alpha\mu\nu}$, $j^\nu=q \int ds u^\nu \delta(x-x(s))\dfrac{1}{\sqrt{-g}}$ is  a four-current of relativistic particles if $q$ is considered as an electric charge,
 $ T_{1\mu \nu}+T_{2\mu\nu}=\rho_m u_\mu u_\nu - \frac{1}{4\pi} (F_{\mu \alpha} F_\nu{}^\alpha-\frac{1}{4}g_{\mu\nu}F^{\alpha\beta}F_{\alpha\beta})$ is the stress-energy tensor of a relativistic particle and electromagnetic field, where
 $\rho_m =m \int ds \delta(x-x(s))\frac{1}{\sqrt{-g}}$ is a mass density and 
$u^\mu=\dot{x}^\mu \frac{1}{\sqrt{\dot{x}^\alpha\dot{x}^\beta g_{\alpha\beta}}}$ is a four-velocity. The role of electromagnetic potential is played by trace part of connection $\hat{\Gamma}^\alpha_{\mu\nu}$, i.e field $\hat{B}_\mu$. Thus the modification of gravity with the action of the gravitational field $S_1+S_2$ and interaction with particles in the form $S_{int}$  turns out to be completely equivalent to the Einstein-Maxwell theory with matter in the form of a pointlike particles.

\subsection{Scalar field}
From the modern point of view, more interesting case is
a theory with a matter in form of fields. If one add a scalar field $\varphi$ in such theory it is necessary to change terms $S_m$ and $S_{int}$ for $S_{\varphi}$:
\begin{equation}
 S_\varphi=\int d^4x \sqrt{-g}((\partial_\mu+i\hat{B}_\mu)\varphi)^\ast(\partial_\nu+i\hat{B}_\nu)\varphi)g^{\mu\nu}.
\end{equation}
We choose this form of action because it looks like the one in a standard approach, but there is a field $\hat{B}_\nu$ instead of electromagnetic potential. Hence in this case the total action consists of three terms: $S_1+S_2+S_{\varphi}$. This theory depends on three independent variables: the metric $g_{\mu\nu}$, 
 the connection $\Gamma^\mu_{\alpha\gamma}$ and scalar field $\varphi$. 

The equations of motion again reproduce the Einstein-Maxwell theory. The quantity  $\tilde{\Gamma}^\mu_{\alpha\gamma}$ is equal to zero again due to the equations of motion.

\subsection{Spinor field}
If we want to consider a spinor field, we need to introduce new objects.
First of all instead of metric it is necessary to consider a  tetrad $e^a_\mu$, which is related to the metric by  relation
\begin{equation}
g_{\mu\nu}=e^a_\mu e^b_\nu \eta_{ab},
\end{equation}
where $ \eta_{ab}$ is a flat metric. 
Also one need to define a Riemannian spin-connection $\bar{\omega}_\mu{}^a{}_b$ from  the condition that  the Riemann covariant derivative of the tetrad is equal to zero:
\begin{gather}
 \bar{D}_\mu e^a_\nu= \partial_\mu e^a_\nu -\bar{\Gamma}^\alpha_{\mu\nu} e^a_\alpha +\bar{\omega}_\mu{}^a{}_b e^b_\nu=0 \longrightarrow \bar{\omega}_\mu{}^a{}_b=\bar{\Gamma}^\alpha_{\mu\nu}e^a_\alpha e^\nu_b -e^\nu_b\partial_\mu e^a_\nu.
\end{gather}
Now there are two  connections: coordinate connection $\Gamma^\alpha_{\mu\nu}$ and spin-connection $\bar{\omega}_\mu{}^a{}_b$. The last one depends on Christoffel symbols, e.g. on metric, which  contains tetrads. Therefore $\Gamma^\alpha_{\mu\nu}$ and $\bar{\omega}_\mu{}^a{}_b$ are not related to each other. We can proceed similarly to what was done above for a scalar field. Namely, one can  write the standard action for the spinor field, which interacts with spin-connection  $\bar{\omega}_\mu{}^a{}_b$ and electromagnetic potential, but replacing the field $\hat{B}_\nu$ in a place of electromagnetic potential. As a result the quantity  $\tilde{\Gamma}^\mu_{\alpha\gamma}$ will be equal to zero again and we get a trivial unification of gravity and electromagnetism.

However, one can consider a more natural unification.
Namely, let us extended local symmetry group from
$SO(1,3) \approx SL(2,C)$ to $ GL(2,C)$. For the simplicity, let us consider Weyl spinor $\chi^A$ (here and below $A=1,2$) instead of Dirac spinor. We use  $(\chi^A)^\ast\equiv \chi^{\dot{A}}$ for a complex conjugate spinor. To consider $GL(2,C)$ local group we need to introduce  independent tetrad in this form
\begin{gather}
 e_{A\dot{B}}^\mu=\frac{1}{\sqrt{2}}\sigma^a_{A\dot{B}}e_a^\mu;\,\,\,\,\,\,\,\,
 (e^\mu_{A\dot{B}})^\ast =e^\mu_{B\dot{A}}.
\end{gather}
There is a problem from very beginning, because group $GL(2,C)$ does not contain any universal tensor to construct a metric tensor.  Therefore we need  to replace the antisymmetric quantity $\varepsilon_{AB}$ by new 
arbitrary antisymmetric field
$E_{AB}$, which must be proportional to $\varepsilon_{AB}$:
\begin{equation}
E_{AB}(x)=\varphi(x)\varepsilon_{AB},
\end{equation}
where $\varphi(x)$ is some nondegenerate field.
Therefore metric has this form
\begin{equation}
g_{\mu\nu}=e_\mu^{A\dot{B}}e_\nu^{C\dot{D}}E_{AC}E_{\dot{B}\dot{D}}.
\end{equation}

From the geometric reasons connection $\Gamma^\alpha_{\mu\nu}$ can be  related with independent $GL(2,C)$ spin-connection $\omega_\mu{}^A{}_B$ by  the condition that  the general covariant derivative of the tetrad is equal to zero:
\begin{equation}
D_\mu e_\nu^{A\dot{B}}=\partial_\mu e_\nu^{A\dot{B}}-\Gamma^\sigma_{\mu\nu}e_\sigma^{A\dot{B}}+
\omega_\mu{}^A{}_Ce_\nu^{C\dot{B}}+\omega_\mu{}^{\dot{B}}{}_{\dot{D}}e_\nu^{A\dot{D}}=0.
\end{equation}
So the expression for connection is following
\begin{equation}
 \Gamma^\alpha_{\mu\nu}(\omega)=e^\alpha_{A\dot{B}}(\partial_\mu e_\nu^{A\dot{B}}+\omega_\mu{}^A{}_C
e_\nu^{C\dot{B}}+\omega_\mu{}^{\dot{B}}{}_{\dot{D}}e_\nu^{A\dot{D}}).
\end{equation}
Therefore now independent variables are tetrad $e_{A\dot{B}}^\mu$, field $\varphi(x)$, spinor $\chi^A$ and
 spin-connection  $\omega_\mu{}^A{}_C$ instead of connection $\Gamma^\alpha_{\mu\nu}$ . 
 
 We use the Einstein-Hilbert action (1) as the basic action of the theory. Instead of (10) let us consider action
 \begin{equation}
\tilde{S}_2=-\theta \int d^4x\sqrt{-g} g^{\alpha\mu}g^{\beta\nu}F^A{}_{A\alpha
\beta}F^B{}_{B\mu\nu},  
\end{equation}
where $F^A{}_{B\mu\nu}=[\partial_\mu \omega_\nu{}^A{}_B+\omega_\mu{}^A{}_C
\omega_\nu{}^C{}_B]_{\mu\nu}$ is a standard tension tensor, but depends on $GL(2,C)$ spin-connection.  
The role of the action of matter is played by standard  action for Weyl spinor 
\begin{equation}
 S_\chi=\int d^4x \sqrt{-g}(i\chi^{\dot{A}} e^\mu_{B\dot{A}}D_\mu \chi^B-i\chi^A e^\mu_{A\dot{B}}D_\mu \chi^{\dot{B}}),
\end{equation}
where $D_\mu \chi^B=(\delta^A_C\partial_\mu+\omega_\mu{}^A{}_C)\chi^C$ consist of general $GL(2,C)$ 
spin-connection $\omega_\mu{}^A{}_C$.
 If one make this change of variables:
 \begin{gather}
  e^{A\dot{B}}_\mu \longrightarrow \tilde{e}^{A\dot{B}}_\mu=e^{A\dot{B}}_\mu |\varphi|;\\
 \chi^A \longrightarrow \tilde{\chi}^A=\chi^A\sqrt{|\varphi|},
 \end{gather}
then field $\varphi$ is fell out from the action at all. This means that we have a possibility not to consider the field $\varphi$  as an independent variable.

As we did it above for the connection $\Gamma^\alpha_{\mu\nu}$  let us split spin-connection to the traceless part and
imaginary and real trace parts
\begin{equation}
 \omega_\mu{}^A{}_B=\underbrace{\tilde{\omega}_\mu{}^A{}_B}_{\tilde{\omega}_\mu{}^A{}_A=0}+\frac{1}{2}(iA_\mu+B_\mu)\delta^A_B.
\end{equation}
Using this notation it is easy to find quantity $F^A{}_{A\mu\nu}$ from (21):
\begin{equation}
 F^A{}_{A\mu\nu}=2(\partial_\mu(B_\nu+iA_\nu)-\partial_\nu(B_\mu+iA_\mu)).
\end{equation}

Let us define  $\hat{\omega}_\mu{}^A{}_B$ which is
equal to the difference between traceless part of $GL(2,C)$ connection and Riemannian spin-connection:
  \begin{equation}
   \hat{\omega}_\mu{}^A{}_B=\tilde{\omega}_\mu{}^A{}_B-\bar{\omega}_\mu{}^A{}_B,
  \end{equation}
where  Riemannian spin-connection $\bar{\omega}_\mu{}^A{}_B$ is traceless and defined from this relation:
 \begin{equation}
  \partial_\mu \tilde{e}_\nu^{A\dot{B}}-
  \bar{\Gamma}^\alpha_{\mu\nu}\tilde{e}_\alpha^{A\dot{B}}+\bar{\omega}_\mu{}^{\dot{A}}{}_{\dot{C}}
  \tilde{e}_\nu^{B\dot{C}}+\bar{\omega}_\mu{}^B{}_D \tilde{e}_\nu^{D\dot{A}}=0.
 \end{equation}
Spin-connection $\omega_\mu{}^A{}_B$  can be written in the following form, using the notations introduced above
\begin{gather}
 \omega_\mu{}^A{}_B=\bar{\omega}_\mu{}^A{}_B+ \hat{\omega}_\mu{}^A{}_B+\frac{1}{2}(iA_\mu+B_\mu)\delta^A_B.
 \end{gather}
 It is useful decomposition, because we can consider separately $\hat{\omega}_\mu{}^A{}_B$, $A_\mu$ and $B_\mu$ as independent variables instead of more complicated $\omega_\mu{}^A{}_B$.
 Also one can find relation between traceless part of connection $\tilde{\Gamma}_{\alpha\mu\nu}$ and $ \hat{\omega}_\mu{}^A{}_B$, using (19) and (27):
 \begin{gather}
 \tilde{\Gamma}_{\alpha\mu\nu}=2 Re (\hat{\omega}_{\mu}{}^F{}_ {D}\tilde{e}_\alpha^{B\dot{C}}\tilde{e}_\nu^{D\dot{A}}\varepsilon_{\dot{A}\dot{C}}\varepsilon_{FB}).
\end{gather}
From this relation and using traceless of $\hat{\omega}_{\mu}{}^F{}_ {D}$, one can find the antisymmetry property: $\tilde{\Gamma}_{\alpha\mu\nu}=-\tilde{\Gamma}_{\nu\mu\alpha}$.
Hence $\tilde{\Gamma}_{\alpha\mu\nu}$ and $ \hat{\omega}_\mu{}^A{}_B$  have the same number of degrees of freedom, therefore they are uniquely connected with each other. Thus $\tilde{\Gamma}_{\alpha\mu\nu}$  can be considered as an independent quantity instead of  $\hat{\omega}_\mu{}^A{}_B$.

Let us get back to the action, which consists of three terms: $S=S_1+\tilde{S}_2+S_\chi$. Using the notations introduced above (see equations (7), (22)-(26), (29)), one can rewrite it in the following form
 \begin{multline}
   S=-\frac{1}{2\varkappa} \int d^4x \sqrt{-g} g^{\mu\nu} \bar{R}_{\mu\nu}-\\ -\theta \int d^4x \sqrt{-g} g^{\alpha\rho}g^{\beta\sigma}
((\partial_\alpha A_\beta-\partial_\beta A_\alpha)(\partial_\rho A_\sigma-
\partial_\sigma A_\rho)+(\partial_\alpha B_\beta-\partial_\beta B_\alpha)(\partial_\rho B_\sigma-
\partial_\sigma B_\rho))+\\+\int d^4x \sqrt{-g}(\tilde{\chi}^{\dot{A}} i 
\tilde{e}^\mu_{B\dot{A}}(\delta^B_C\partial_\mu+\bar{\omega}_\mu{}^B{}_C+
iA_\mu \delta^B_C)\tilde{\chi}^C+ h.c.) -\\-\frac{1}{4\varkappa} \int d^4x \sqrt{-g}
 \tilde{\Gamma}_{\mu\alpha\gamma}\tilde{\Gamma}_{\delta\beta\nu} \tilde{L}^{\mu\alpha\gamma,\delta\beta\nu}
+\frac{1}{2}\int d^4x\sqrt{-g}E_{\beta\alpha\nu\mu}\tilde{\Gamma}^{\alpha\nu\mu}
\tilde{\chi}^{\dot{A}}\tilde{e}^\beta_{B\dot{A}}\tilde{\chi}^B,
 \end{multline}
 where $\tilde{L}^{\mu\alpha\gamma,\delta\beta\nu}$ is an antisymmetric part of  $L^{\mu\alpha\gamma,\delta\beta\nu}$ on the $\mu, \gamma$ and $\delta, \nu$ indices. We consider $\tilde{L}^{\mu\alpha\gamma,\delta\beta\nu}$ as $24\times 24$ matrix, which is acting on 24 components of $\tilde{\Gamma}_{\delta\beta\nu}$ (taking into account the antisymmetry property of  $\tilde{\Gamma}_{\delta\beta\nu}$).
The connection $\tilde{\Gamma}^{\alpha\nu\mu}$ is contained only in the last two terms. If one varies the action with respect to $\tilde{\Gamma}^{\alpha\nu\mu}$,  the following equation is obtained
\begin{equation}
   \tilde{\Gamma}_{\mu\alpha\gamma}=\varkappa \tilde{L}^{\text{-1}}_{\mu\alpha\gamma,\delta\beta\nu}E^{\sigma\delta\beta\nu}j_\sigma,
\end{equation}
where $j^\beta=\tilde{\chi}^{\dot{A}}\tilde{e}^\beta_{B\dot{A}}\tilde{\chi}^B$ and $\tilde{L}^{\text{-1}}_{\mu\alpha\gamma,\delta\beta\nu}$ is inverse matrix of $\tilde{L}^{\mu\alpha\gamma,\delta\beta\nu}$.
 Substituting this back into the last two terms, one can see that they are equal to zero. It is happened due to the  fact that terms are proportional to square of current $j^\beta$, which is zero because of involution symmetric with antisymmetric:
\begin{equation}
 j^\alpha j^\beta g_{\alpha\beta}=\tilde{\chi}^{\dot{A}}\tilde{e}^\alpha_{B\dot{A}}\tilde{\chi}^B
 \tilde{\chi}^{\dot{C}}\tilde{e}^\beta_{D\dot{C}}\tilde{\chi}^D g_{\alpha\beta}=\tilde{\chi}^{\dot{A}}\tilde{\chi}^B
 \tilde{\chi}^{\dot{C}}\tilde{\chi}^D \varepsilon_{BD}\varepsilon_{\dot{A}\dot{C}}=0.
\end{equation}
Thereby the total action fully coincides with the action in standard Einstein-Maxwell-Dirac approach, with the exception of term with massless  field $B_\mu$. This field interacts only with gravity and we can interpret it as dark photon. The role of electromagnetic potential  is played by $A_\mu$ which is imaginary part of trace $GL(2,C)$ spin-connection. 
Thus, within the framework of this approach, a theory equivalent to Einstein-Maxwell-Dirac was obtained.
It is important that the square of current $j^\beta$ is vanished only for Weyl spinor. If we consider Dirac spinor, the last two terms from (30) will give contributions to the motion equations  proportional to $\varkappa$.

\section{Acknowledgments}
This research is supported by a grant from the Russian Foundation for Basic Research (Project No. 18-31-00169).

\end{document}